\documentclass[10pt,twocolumn,showpacs,preprintnumbers,amsmath,amssymb,prb,superscriptaddress,longbibliography]{revtex4-2}
\usepackage{bbm}
\usepackage{mathrsfs}
\usepackage{graphicx}
\usepackage{dcolumn}
\usepackage{bm}
\usepackage{braket}
\usepackage{amsmath}
\usepackage{amsfonts}
\usepackage[dvipsnames]{xcolor}
\usepackage[colorlinks=true,linkcolor=Blue,urlcolor=BlueViolet,citecolor=BlueViolet]{hyperref}
\usepackage{floatrow}
\usepackage{multirow}
\begin{document}

\title{Josephson diode effect in topological superconductor}
\author{Zhaochen Liu}
\affiliation{State Key Laboratory of Surface Physics and Department of Physics, Fudan University, Shanghai 200433, China}
\affiliation{Shanghai Research Center for Quantum Sciences, Shanghai 201315, China}
\author{Linghao Huang}
\affiliation{State Key Laboratory of Surface Physics and Department of Physics, Fudan University, Shanghai 200433, China} 
\affiliation{Shanghai Research Center for Quantum Sciences, Shanghai 201315, China}
\author{Jing Wang}
\thanks{wjingphys@fudan.edu.cn}
\affiliation{State Key Laboratory of Surface Physics and Department of Physics, Fudan University, Shanghai 200433, China}
\affiliation{Shanghai Research Center for Quantum Sciences, Shanghai 201315, China}
\affiliation{Institute for Nanoelectronic Devices and Quantum Computing, Fudan University, Shanghai 200433, China}
\affiliation{Hefei National Laboratory, Hefei 230088, China}

\begin{abstract} 
We investigate the Josephson diode effect (JDE) in topological Josephson junctions. By both analytic and numerical calculations, we find that while a Josephson junction in the topological phase may exhibit higher diode efficiency compared to that in the trivial phase, this behavior is not universal. The presence of Majorana bound states is not a sufficient condition for a large diode effect. Furthermore, the diode efficiency undergoes substantial changes only in \emph{specific} regions along the topological phase transition boundary, and a significant diode effect does coincide with the topological phases. Thereby our paper suggests the utilization of topological superconductivity for enhanced JDE, and also the Josephson diode effect may serve as an indicator for topological superconductor phase. These results suggest a nuanced relationship between the topological aspects of Josephson junctions and Josephson diode effect.
\end{abstract}


\maketitle

\section{Introduction} 
Superconductivity, a fascinating macroscopic quantum phenomenon with profound physical significance and diverse practical applications, has been the focus of extensive research for decades~\cite{tinkham2004}.
In recent years, the burgeoning field of topological superconductivity has emerged as an exciting frontier in condensed matter physics~\cite{qi2011topological,tanaka2012symmetry,ando2015,chiu2016,sato2017}.
Topological superconductors are predicted to host exotic quasiparticles, Majorana bound states (MBS)~\cite{kitaev2001unpaired,leijnse2012introduction,beenakker2013search}, which are topologically protected and exhibit non-Abelian exchange statistics~\cite{ivanov2001nonabelian}, and have potential applications in topological quantum computation~\cite{kitaev2003faulttolerant,nayak2008nonabeliana,aasen2016milestones}. One of the most promising avenues for realizing topological superconductivity is through the Josephson junctions~\cite{fu2008superconducting,lutchyn2010majorana,oreg2010helical,alicea2010majorana,hell2017twodimensional,pientka2017topological}, which are composed of two superconductors interconnected by a weak-link. When a junction is expected to transition into a topological superconducting phase, unpaired MBS will emerge at the junction interfaces.
So far substantial theoretical studies have investigated the exotic properties that MBS may exhibit in the topological Josephson junctions~\cite{tanaka1996theory,tanaka1997theory,tanaka2009manipulation,tanaka2010anomalous,san-jose2012ac,tanaka2012symmetry,cayao2015sns,dolcini2015topological,peng2016signatures,cayao2017majorana,cayao2018andreev,chen2018asymmetric,tokura2018nonreciprocal,cayao2021distinguishing,baldo2023zerofrequency,dartiailh2021phase,tanaka2021theory}.

Recently, a diode effect has been proposed and discovered in Josephson junctions, termed as the Josephson diode effect (JDE)~\cite{hu2007proposed,misaki2021theory,davydova2022universal,wang2022symmetry,zhang2022general, baumgartner2022supercurrent,wu2022fieldfree,pal2022josephson,liu2013,nagaosa2024nonreciprocal}.
A Josephson diode has asymmetric critical currents in the forward ($I^{+}_c$) and backward ($I^{-}_c$) directions~\cite{nadeem2023superconducting}. Therefore, it manifests superconductive in one direction while resistive in the other direction when the magnitude of current falls within the range $\text{min}\{I^{-}_c,I^{+}_c\}<I<\text{max}\{I^{-}_c,I^{+}_c\}$. In comparison to superconducting diode effect in bulk systems~\cite{yuan2021topological,daido2022intrinsic,yuan2022supercurrent,he2022phenomenological,takasan2022supercurrentinduced,zinkl2022symmetry,ando2020observation,miyasaka2021observation,narita2022fieldfree,hou2023ubiquitousa}, Josephson junctions may attain higher diode efficiency due to the suppression of kinetic energy, and thereby the enhancement of the influence of interactions~\cite{tanaka2022theory}. Additionally, the supercurrent within a Josephson junction can be more controllable by adjusting the phase difference between the two superconductors~\cite{likharev1979superconducting}, facilitating a deeper understanding of the underlying mechanism in diode effect.

Since the appearance of MBS in topological superconductors induces alterations in the energy spectrum of Josephson junctions, it further modifies the current-phase relation. This prompts the consideration of the potential impact of MBS on the JDE. However up to date, there is less focus on the feature of JDE in the presence of MBS~\cite{nesterov2016anomalous,spanslatt2018geometric,kutlin2020geometrydependent,kopasov2021geometry,tanaka2022theory,murthy2020energy,legg2023parity,cuozzo2024microwavetunable}. These studies provide some insights into the  relationship between diode effect and topological phase transitions, but the influence of MBS on JDE as well as the relation between topological superconductors and JDE still needs more investigations.

Here we study the JDE in two representative systems consisting of topological superconductors. Specifically, we consider a proximitized semiconducting nanowire with strong Rashba spin-orbital coupling (SOC), and a two-dimensional magnetic topological insulator (TI) thin film proximity coupled to $s$-wave superconductivity. By employing the Bogoliubov-de Gennes (BdG) mean-field calculation, we find that the presence of MBS in the topological phase strongly affects the asymmetry of the Andreev spectrum, which leads to JDE. Moreover, the topological phase can exhibit higher diode efficiency compared to the trivial phase under certain conditions, specifically the diode efficiency undergoes substantial changes only in specific regions along the topological phase transition boundary, and a significant diode effect does coincide with the topological phases. Our paper suggests the utilization of topological superconductivity for enhanced JDE, while it also suggests JDE may serve as an indicator for topological superconductors in experiments.

\section{Rashba Nanowire}
\subsection{Model}
\begin{figure}[t] 
  \begin{center}
  \includegraphics[width=3.4in,clip=true]{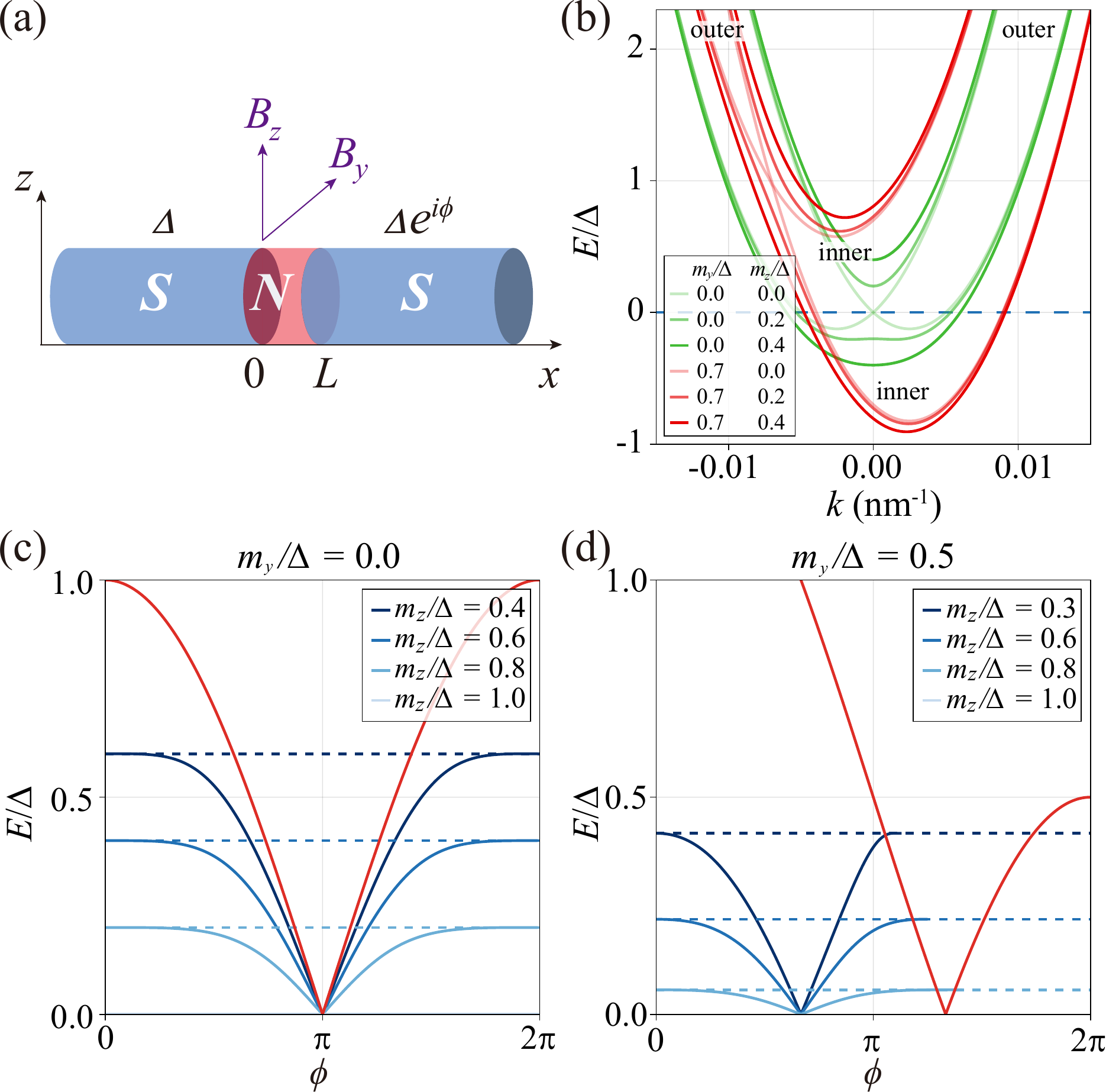}
  \end{center} 
  \caption{(a) Sketch of a Josephson junction along the $x$-axis with normal state region of finite length $L$. The SOC is along the $y$-axis and external magnetic field has both $y$ and $z$ components. (b) Band structure of normal state in Rashba nanowire. $m_z$ term opens an energy gap while $m_y$ asymmetrizes the band structure. The system consists of a pair of outer modes and a pair of inner modes, and the induced $s$-wave superconducting pairing acts within each pair of modes. (c) and (d) The ABS spectrum with respect to different $m_z$ for $m_y/\Delta=0$ and $m_y/\Delta=0.5$, respectively. The red (blue) lines are spectrum of outer (inner) modes, and the dashed lines mark the upper bound energy for inner modes.}
  \label{fig1}
\end{figure}

We start from a paradigmatic system of Josephson junctions: a Rashba nanowire with proximity-induced pairing correlation, see Fig.~{\ref{fig1}}(a). The BdG Hamiltonian for this system is given by~\cite{lutchyn2010majorana,oreg2010helical}
\begin{equation}\label{RashbaHamiltonian}
    \hat{H}_{\text{1d}}=\frac{1}{2}\int dx \hat{\Psi}^{\dagger}(x) \mathcal{H}_{\text{1d}}\hat{\Psi}(x)
\end{equation}
with $\hat{\Psi}(x)\equiv(\psi_{\uparrow},\psi_{\downarrow},\psi^{\dagger}_{\uparrow},\psi^{\dagger}_{\downarrow})^T$, and
\begin{equation} 
    \mathcal{H}_{\text{1d}}=\left(\frac{k^2}{2m}-\mu+m_z\sigma_z\right)\tau_z+m_y\sigma_y+\alpha k \sigma_y\tau_z+\Delta(x),\nonumber
\end{equation}
where $k\equiv-i\hbar\partial_x$, the Zeeman coupling $m_y$ and $m_z$ are induced by the external magnetic field, $\mu$ is the chemical potential, $\alpha$ is the SOC strength and $\Delta(x)$ is the $s$-wave pairing potential. $\sigma_i$ and $\tau_i$ ($i=x,y,z$) are Pauli matrices for spin and Nambu space, respectively. For the Josephson junction with phase difference $\phi$ between two superconductors, the pairing is $\Delta(x)=\Delta\sigma_y\tau_y$ for $x\leq 0$ and $\Delta(x)=\Delta e^{i\phi}\sigma_y(\tau_y-i\tau_x)/2+\text{h.c.}$ for $x>L$, with $L$ as the length of the junction. We focus on the experimentally important case of short junction.

In general, the Josephson current can be divided into two parts: one comes from the Andreev bound states (ABS) and the other is contributed from the continuum states~\cite{san-jose2013multiple,kopasov2021geometry,davydova2022universal}. The Josephson current is related to the spectrum of $\mathcal{H}_{\text{1d}}$ by~\cite{beenakker1991universal} 
\begin{equation}\label{currentEnergyRelation}
 I(\phi)=\frac{2e}{\hbar}\frac{\partial E}{\partial \phi}, \quad E=-\frac{1}{2}\sum_{E_n\geq0} E_n(\phi),
\end{equation}
where the summation includes both the ABS and continuum states. Hereafter, we set $e=1$ and $\hbar=1$.

\subsection{Symmetry analysis}
Before explicit calculation, we first analyze the role of symmetries on the critical current $I_{c}^{\pm}$~\cite{zinkl2022symmetry,wang2022symmetry,zhang2022general}, denoting the maximum amplitude of the Josephson current for forward and backward directions, respectively. Without loss of generality, we assume that the Josephson current flows along the $x$ axis, then the operations of time reversal ($\mathcal{T}$), mirror reflection with respect to the $x$ axis ($\mathcal{M}_x$), and space inversion ($\mathcal{P}$) can reverse the current direction, whereas the other two mirror reflection operations ($\mathcal{M}_y$ and $\mathcal{M}_z$) can not. Taking $\mathcal{T}$ as an example, if a system exhibits $\mathcal{T}$ symmetry when no Josephson current flows across the junction (with its Hamiltonian denoted as $H(0)$), then the system with forward current $H(I_0 \hat{x})$ is related to the system with backward current $H(-I_0 \hat{x})$ via $\mathcal{T}$ operation. Therefore, the forward and backward critical current are equal: $I_{c}^{+}=I_{c}^{-}$, indicating the absence of a diode effect. Similarly, $\mathcal{M}_x$ ($\mathcal{P}$) also ensure $I_{c}^{+}=I_{c}^{-}$. Conversely, $\mathcal{M}_y$ and $\mathcal{M}_z$ symmetries impose no constraints on $I_{c}^{\pm}$. Hence symmetries like $\mathcal{T}$, $\mathcal{M}_x$, $\mathcal{P}$, or combined symmetries such as $\mathcal{TM}_{y}$, $\mathcal{TM}_{z}$, $\mathcal{M}_{x}\mathcal{M}_{y}$, $\mathcal{M}_{x}\mathcal{M}_{z}$, $\mathcal{PM}_{y}$, $\mathcal{PM}_{z}$ and $\mathcal{TM}_{y}\mathcal{M}_{z}$ can enforce the vanishing of JDE. In the system under consideration, $\mathcal{T}=-i\sigma_y\mathcal{K}$ where $\mathcal{K}$ is complex conjugate, $\mathcal{M}_x=\sigma_x$, $\mathcal{M}_y=\sigma_y$, $\mathcal{M}_z=\sigma_z$ and $\mathcal{P}=\sigma_0 \tau_0$. The $\mathcal{P}$ symmetry requires $\mathcal{P} H(k) \mathcal{P}^{-1} = H(-k)$, which has been broken by the SOC term, $\alpha k \sigma_y \tau_z$. Without the Zeeman term, the system exhibits $\mathcal{T}$, $\mathcal{M}_x$, $\mathcal{TM}_{y}$, $\mathcal{M}_{x}\mathcal{M}_{y}$ and $\mathcal{PM}_{z}$ symmetries, thus has no diode effect. If $m_z\neq 0$ and $m_y=0$, $\mathcal{T}$ and $\mathcal{M}_x$ are broken but there remain $\mathcal{TM}_{y}$, $\mathcal{M}_{x}\mathcal{M}_{y}$ and $\mathcal{PM}_{z}$, giving rise to a zero diode effect. For $m_z=0$ and $m_y\neq 0$, all of the above symmetries are broken. However, there is a hidden symmetry with $U(x)\mathcal{P}U(x)^\dagger$, where $U(x)=\exp(-i\alpha m x\sigma_y/2)$ is the spin twist operator~\cite{wang2022symmetry}. $U(x)$ eliminates SOC, and $\mathcal{P}$ symmetry requires $I_{c}^{+}=I_{c}^{-}$. Therefore, to obtain JDE, both the Zeeman terms $m_y$ and $m_z$ are necessary.

As a side note, in this system, zero supercurrent is reached at $\phi=0$. If an anomalous phase shift occurs, zero supercurrent will arise at a finite phase difference $\phi=\phi_0$. Therefore, the symmetry analysis should be done at $\phi-\phi_0=0$, and the property of $\phi_0$ under symmetry transformations should be considered in the symmetry analysis. However, the symmetry requirement itself does not change; all the symmetries mentioned above should still be broken.

\subsection{Results}
In what follows, we discuss the JDE and topological phase transition in this system by analyzing the Andreev spectrum. The ABS can be determined using the wave function matching condition in the short junction limit (i.e., junction length $L~\ll\xi$ superconducting coherence length). For simplicity of analytical calculation, we first consider the short junction limit with $L=0$ and choose $\mu=0$. Assuming that SOC dominates  $m\alpha^2\gg\Delta,m_y,m_z$, then the linearized Hamiltonian for the low-energy physics of inner and outer modes [Fig.~\ref{fig1}(b)] can be obtained near $k_\text{in}=0$ and $k_\text{out}=\pm 2m\alpha$, respectively~\cite{vanheck2017zeeman,murthy2020energy}
\begin{equation}\label{linearizedHamiltonian}
\begin{aligned}
H_\text{in}&=\alpha k\sigma_y\tau_z+m_z\sigma_z\tau_z+m_y\sigma_y\tau_0+\Delta(x),
\\
H_\text{out}&=-\alpha k\sigma_y\tau_z +m_y\sigma_y\tau_0+\Delta(x).
\end{aligned}
\end{equation}
For $H_\text{out}$, $\sigma_y=\mp$ corresponds to $k_\text{out}=\pm 2m\alpha$ modes. We assume that the inter-mode scattering at the junction interface is neglected, and thus treat different modes independently. The Hamiltonian with $m_z=0$ resembles pervious results on the finite momentum pairing~\cite{davydova2022universal}, which can be seen from the gauge transformation $\psi(x)\rightarrow \exp(\mp im_yx/\alpha)\tau_z\psi(x)$ for inner and outer modes, respectively. Then the $m_y$ term is eliminated, while the pairing term carries an additional phase factor $e^{\mp i2m_y x / \alpha}$, which means Cooper pairs in two modes get opposite momentum. Then the Andreev spectrum of outer modes are
\begin{equation}\label{outDispersion}
    E^\text{out}_1=\Delta \cos(\phi/2)+m_y, E^\text{out}_{2}= -\Delta\cos(\phi/2)-m_y.
\end{equation}
The current-phase relation is obtained from Eq.~(\ref{currentEnergyRelation}) as $I_\text{out}=\Delta \sin(\phi/2) \text{sgn}(\Delta \cos(\phi/2)+m_y)/2$. Thus the contribution from ABS to JDE is $I_{c}^{+}-I_{c}^{-}=\text{sgn}(m_y)\Delta(1-\sqrt{1-(m_y/\Delta)^2})/2$. For inner modes, when $m_z=0$, the spectrum is exactly symmetric to the outer modes, which can be seen from the corresponding Hamiltonian and Andreev spectrum
\begin{equation}
    E^\text{in}_{1}=\Delta \cos(\phi/2)-m_y, E^\text{in}_{2}=-\Delta\cos(\phi/2)+m_y.
\end{equation}
This means the contributions of two modes exactly compensate each other. Besides, the Josephson current created by continuum states is independent of $\phi$~\cite{davydova2022universal}. As a consequence, no JDE  occurs, which is consistent with the previous symmetry analysis.

\begin{figure}[t] 
\begin{center}
\includegraphics[width=3.4in,clip=true]{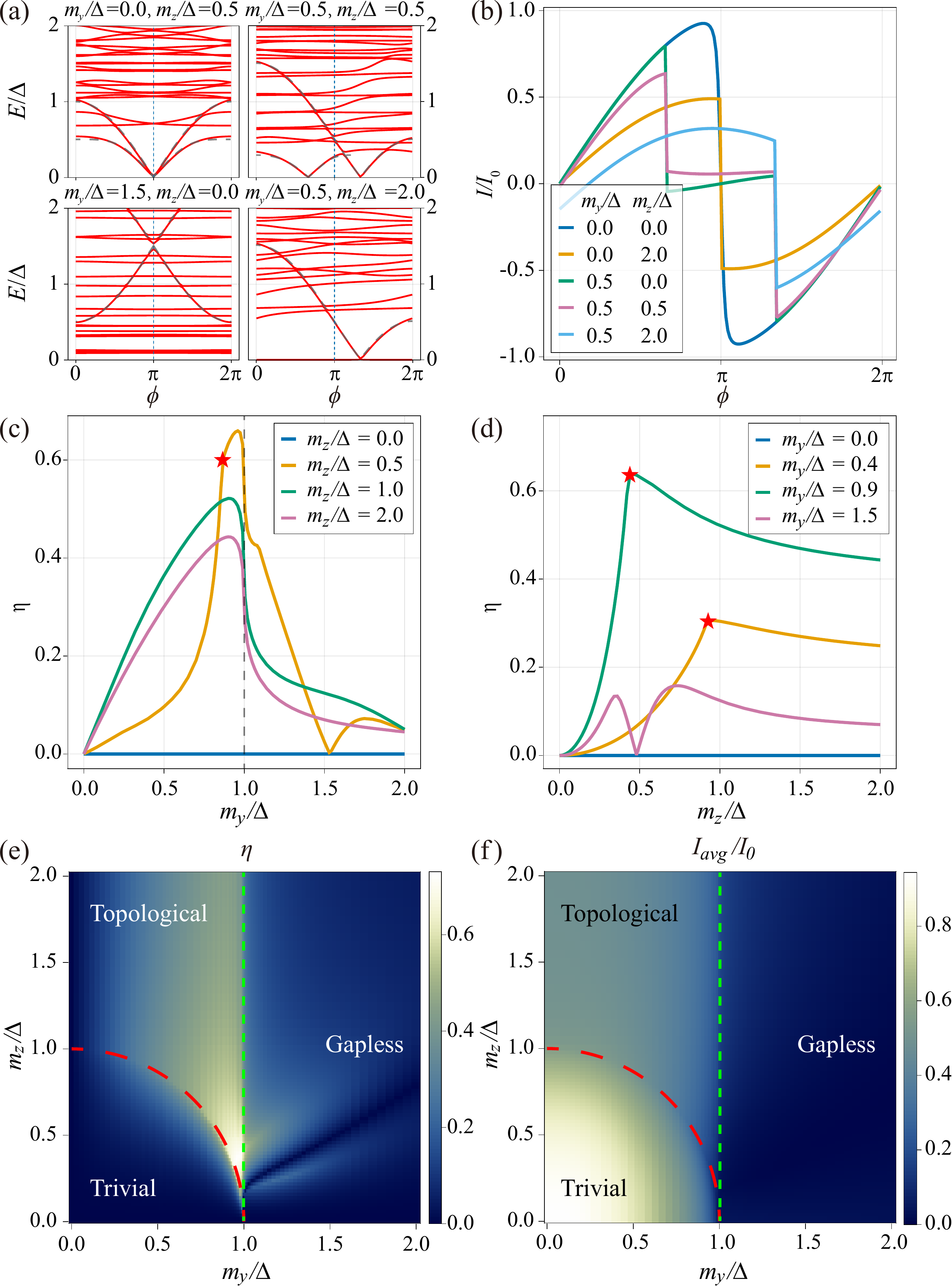}
\end{center} 
\caption{(a) Energy spectrum of one-dimensional short junction near Fermi energy with respect to $\phi$ at different $(m_y, m_z)$. The dashed lines are the analytical results in Eq.~(\ref{outDispersion}). (b) Current-phase relation. (c) $\eta$ vs $m_y$ for typical $m_z$. For $m_z/\Delta<1$, the star symbol marks the boundary between topological and trivial gapped phases. (d) $\eta$ vs $m_z$ for several $m_y$. The star symbol marks the phase boundary when $m_y/\Delta<1$. (e) and (f) The dependence of $\eta$ and $I_\text{avg}$ on $(m_y, m_z)$, where $I_0\equiv e\Delta/\hbar$. The red dashed lines indicate the phase boundary. The green dashed lines delineate the boundary between gapped and gapless regions.}
\label{fig2}
\end{figure} 

Including nonzero $m_z$ does not influence the outer modes, yet the inner modes are no longer symmetric to the outer ones. The corresponding equations become too complicated for analytical solution with nonzero $m_z$ and $m_y$, we solve the spectrum numerically. As shown in Figs.~{\ref{fig1}}(c) and~\ref{fig1}(d), the spectrum is symmetric about $\phi=\pi$ when $m_y\neq 0$, $m_z=0$. With nonzero $m_z$, the inner modes get suppressed, and their contribution to Josephson current would no longer compensate for the outer ones. Thus the imbalance between these two modes generates nonzero JDE. It is expected that the efficiency factor of JDE, $\eta\equiv(|I^{+}_c-I^{-}_c|)/(I^{+}_c+I^{-}_c)$, increases as $m_z$ approaches the topological phase boundary due to further suppression of the inner modes.
Furthermore, we notice that the inner modes are upper bounded by $\Delta-\sqrt{m_z^2+m_y^2}$, so they disappear after the topological transition for $m_z^2+m_y^2>\Delta^2$ and only the outer modes solely contribute to the Josephson current. 

The topology comes from the single Fermi surface condition. In the trivial region, two band inversions occur at the inner and outer Fermi points. While for the topological region, only the band inversion at the outer ones remains. On the other hand, the diode effect is suppressed due to the compensation of these two modes. Thus, coincidence of these two different phenomena could be expected. However, as we can see in the numerical results below, this dose not mean that topology is a \emph{sufficient} condition for a large diode effect.

As studied previously, the contribution from the continuum states is important for determining the magnitude of the asymmetry between the critical currents in opposite directions \cite{davydova2022universal}. In Fig.~{\ref{fig2}}, we numerically calculate the low-energy spectrum, the diode efficiency factor $\eta$, and current-phase relation by taking the contribution from continuum of states into account. We discretize Eq.~(\ref{RashbaHamiltonian}) on the lattice, and set $\Delta=1$~meV, $\alpha=100$~meV$\cdot$nm, $m=2.5\times10^{-3}$~nm$^{-2}\cdot$meV$^{-1}$, $a=10$~\AA, and $L=2a$. Here $m\alpha^2/\Delta=25$ ensures that SOC dominates over superconducting pairing and Zeeman field. We set the total length of nanowire as $1000a$. To get rid of finite size effects, the total DC Josephson current and diode efficiency $\eta$ are calculated by the Green's function method~\cite{furusaki1994dc,zhang2017quantum}. The low-energy spectrum is plotted in Fig.~{\ref{fig2}}(a), where the dashed lines represent the analytical results for the ABS spectrum, which match well with the numerical results. When $m_y=0$, the spectrum is symmetric with respect to $\phi=\pi$, and a similar situation also occurs for $m_z=0$. Only if both $m_y$ and $m_z$ are nonzero, the spectrum can be asymmetric, giving rise to JDE. The current-phase relation for several typical parameters is shown in Fig.~{\ref{fig2}}(b). The discontinuity in the current-phase relation is due to the ABS changing direction at corresponding points.

To better understand JDE, we calculate the evolution of $\eta$ versus $(m_y, m_z)$. For fixed $m_z\neq0$ shown in Fig.~{\ref{fig2}}(c), $\eta$ increases as $m_y$ approaches the phase boundary from trivial side, which accounts for the suppression of ABS from inner modes. When $m_z/\Delta\geq1$, the system is in the topological phase for $m_y<\Delta$ and $\eta$ first increases, and then it enters into gapless phase for $m_y\geq\Delta$ with $\eta$ dramatically decreasing. The $m_y$ term is regarded as the momentum of Cooper pairs under the gauge transformation. Therefore, when $m_y/\Delta<1$, the Doppler shift in energy becomes larger as $m_y$ increases, and the asymmetry between the critical currents becomes larger, resulting in higher $\eta$. The region $m_y/\Delta\geq1$ corresponds to gapless superconductivity, where quasiparticles at zero-energy exist and introduce complexity into the behavior of the Josephson current, and $\eta$ shows nonmonotonic dependence on $m_z$. For fixed nonzero $m_y$ in Fig.~{\ref{fig2}}(d), $\eta$ grows as $m_z$ increases before the phase transition, also resulting from the suppression of inner modes. $\eta$ exhibits a crossover from the trivial to the topological phase. After the phase transition, only outer modes of ABS contribute to Josephson current, and $\eta$ decreases slowly in $m_z$. Interestingly, $\eta$ has a \emph{kink} at the phase boundary between topological and trivial gapped superconductor.

The dependence of $\eta$ and the averaged critical current $I_\text{avg}=|I_{c}^{+}+I_{c}^{-}|$ on $(m_y, m_z)$ are shown in Figs.~{\ref{fig2}}(e) and~\ref{fig2}(f), respectively. We can see that $\eta$ changes significantly along $m_y/\Delta=1$ and part of the phase transition boundary, then JDE seems to be a \emph{weak indicator} for topological phase transition.  $I_\text{avg}$ decreases to be a small value when $m_y>\Delta$, where the critical current $I_{c}^\pm$ is also small, thus JDE in this region has little practical significance. We emphasize that in this model, $\eta$ is continuous but has a kink near the topological phase transition. Thus when $(m_y,m_z)$ are slightly smaller than its critical value for the topological transition, a large $\eta$ can be still reached, where no MBS exists. On the other hand, not all the topological region in Fig.~{\ref{fig2}}(e) exhibits a large $\eta$. These observations suggest that the emergence of MBS may not be the essential factor for enhancing JDE, at least not always. Moreover, when $m_y$ is small, $\eta$ changes little as $m_z$ increases, indicating that in this region, the topological phase transition does not have a significant impact on $\eta$.

\section{Magnetic TI heterostructure}
Now we study JDE in 2D topological superconductivity, which consists of a magnetic TI thin film proximity coupled to an $s$-wave superconductor~\cite{wang2015chiral,wang2016,he2019platform}. The BdG Hamiltonian is described by two surface Dirac fermions with superconducting pairing,
\begin{equation}\label{2Dmodel}
\mathcal{H}_{\text{2d}}=
\begin{pmatrix}
\mathcal{H}_0(\mathbf{k}) & \Delta
\\
\Delta^\dag & -\mathcal{H}^*_0(-\mathbf{k})
\end{pmatrix},
\end{equation}
where $\mathcal{H}_0(\mathbf{k})=v(k_y\sigma_x-k_x\sigma_y)\xi_z+m(\mathbf{k})\xi_x+\mathbf{m}\cdot\mathbf{\sigma}-\mu$, $\Delta=\Delta(x,y)(\xi_0+\xi_z)/2$ is the superconducting pairing induced on the top layer with $\Delta(x,y)=\Delta(x)$. We consider the planar Josephson junction~\cite{chen2018emergent,li2019majoranajosephson,ikegaya2020anomalous,lu2023tunablea} with $\Delta(x)=i\Delta\sigma_y$ for $x\leq0$ and $\Delta(x)=i\Delta e^{i\phi}\sigma_y$ for $x>L$. $\sigma_i$ and $\xi_i$ ($i=x,y,z$) are Pauli matrices for spin and layer, respectively. $v$ is Dirac velocity, $m(\mathbf{k})=m_0+m_1(k_x^2+k_y^2)$ represents the interlayer coupling, $\mathbf{m}$ is the Zeeman term, and $\mu$ is the chemical potential. We focus on the short junction.

We then analyze the symmetry constraints of JDE in this system. Without the Zeeman and pairing terms, the system has time-reversal symmetry $\mathcal{T}=-i\sigma_y\mathcal{K}$, mirror symmetries $\mathcal{M}_x=\sigma_x$, $\mathcal{M}_y=\sigma_y$, $\mathcal{M}_z=\sigma_z\xi_x$ and inversion symmetry $\mathcal{P}=\xi_x$. To obtain a nonzero JDE with Josephson current along the $x$-direction, we need to break symmetries such as $\mathcal{T}$, $\mathcal{M}_x$, $\mathcal{P}$, $\mathcal{TM}_{y}$, $\mathcal{TM}_{z}$, $\mathcal{M}_{x}\mathcal{M}_{y}$, $\mathcal{M}_{x}\mathcal{M}_{z}$, $\mathcal{PM}_{y}$, $\mathcal{PM}_{z}$ and $\mathcal{TM}_{y}\mathcal{M}_{z}$. Therefore, an asymmetric pairing term is introduced to break $\mathcal{P}$, $\mathcal{TM}_{z}$, $\mathcal{M}_{x}\mathcal{M}_{z}$ and $\mathcal{PM}_{y}$. The other symmetries are broken by the Zeeman term $m_y$. Here we consider the external magnetic field is applied along both the $y$ and $z$ axis.

The phase diagram of underlying system has been studied previously~\cite{wang2015chiral}, where the phase boundaries are determined by the bulk BdG gap closing at $\mathbf{k}=0$. For $m_y=0$ and $\mu=0$, the phase boundary is given by $\pm\Delta m_z+m_z^2=m_0^2$. Each gapped phase is characterized by a BdG Chern number $N$. $N=0$ for $|m_z|<(\sqrt{\Delta^2+4m_0^2}-\Delta)/2$, $N=\text{sgn}(m_z)$ when $(\sqrt{\Delta^2+4m_0^2}-\Delta)/2<|m_z|<(\sqrt{\Delta^2+4m_0^2}+\Delta)/2$, and $N=2\text{sgn}(m_z)$ for $|m_z|>(\sqrt{\Delta^2+4m_0^2}+\Delta)/2$. When $m_y\neq0$ and $\mu\neq0$, the phase diagram is obtained through adiabatic evolution to that when $m_y=0$ and $\mu=0$. The $N=0$ and $N=2$ phases are adiabatically connected to trivial insulator and quantum anomalous Hall insulator without gap closure, respectively. $N=1$ phase is the nontrivial topological superconductor which has a single chiral Majorana edge mode, and we will see below the influence of topology on JDE.

\begin{figure}[t] 
\begin{center}
\includegraphics[width=3.4in,clip=true]{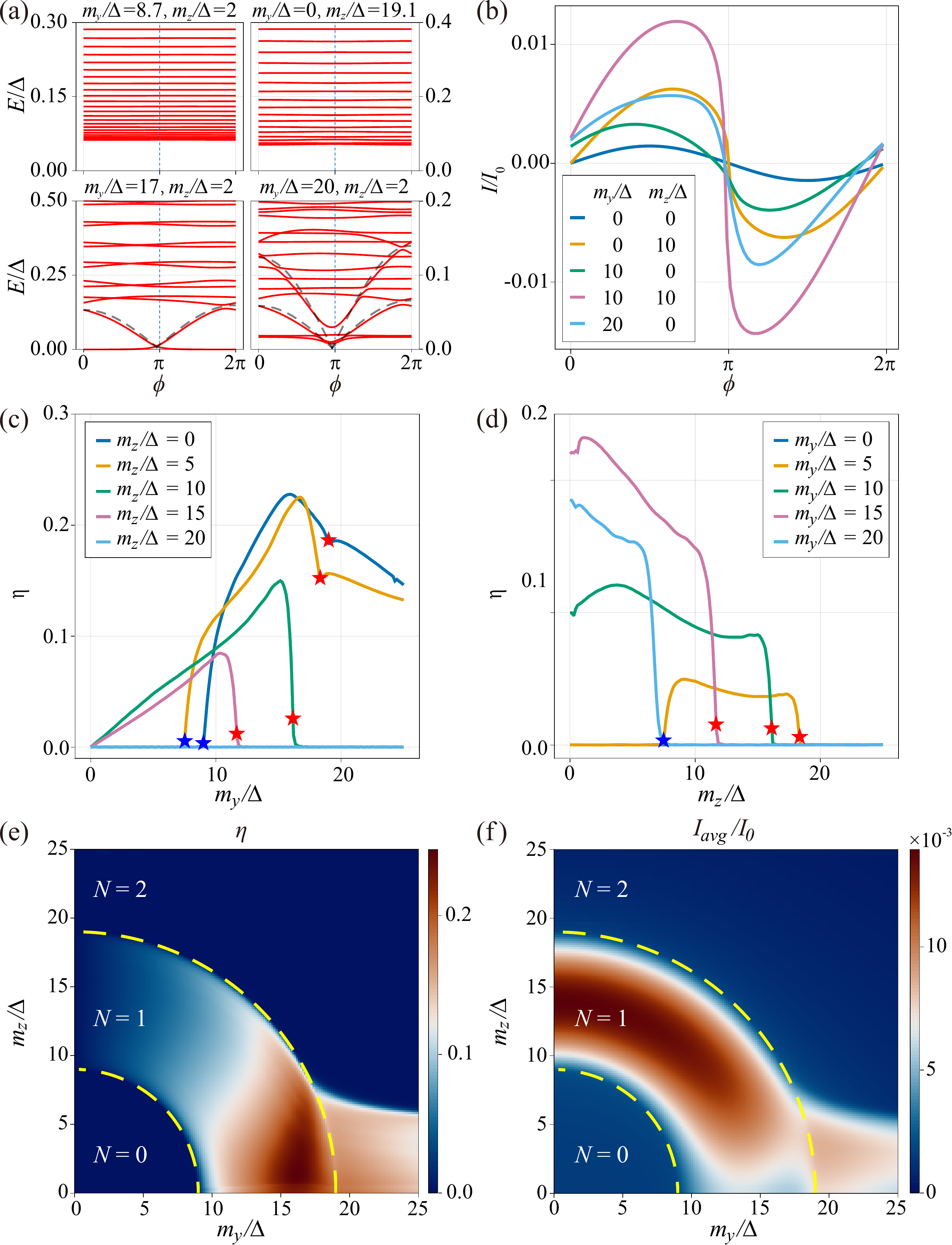}
\end{center} 
\caption{(a) Energy spectrum of 2D magnetic TI heterostructure near Fermi energy with respect to $\phi$ at different $(m_y,m_z)$. (b) Current-phase relation. (c) $\eta$ vs $m_y$. (d) $\eta$ vs $m_z$. The blue star symbol marks the phase boundary between trivial and $N=1$ phase, and the red marks boundary between $N=1$ and $N=2$ phase. (e) and (f) $\eta$ and $I_\text{avg}$ vs $(m_y,m_z)$, respectively. The dashed yellow lines indicate the phase boundary.}
\label{fig3}
\end{figure} 

We numerically calculate JDE in Fig.~\ref{fig3}, and set $m_0=-14$~meV, $a=20~\text{\AA}$, $v=3.2$~eV$\cdot\text{\AA}$, $m_1=9.405$~eV$\cdot{\text{\AA}}^2$, $\Delta=1$~meV, $L=2a$. Furthermore, we set $\mu=5$~meV since a finite chemical potential would enhance the proximity effect and enlarge $N=1$ phase~\cite{qi2010chiral,wang2015chiral}. With periodic boundary condition along the $y$-axis, the Josephson current can be expressed as $I(\phi)=\int dk_y/2\pi ~ I(k_y,\phi)$. In Fig.~{\ref{fig3}}(a), we present the low-energy spectrum of the planar Josephson junction at $k_y=0$ for typical parameters, where the junction length in the $x$-direction is set as $5000a$. For $m_y=0$, the spectrum is symmetric with respect to $\phi=\pi$~\cite{fu2008superconducting,fu2009josephson}. The introduction of a nonzero $m_y$ leads to an asymmetry in the bound state spectrum. Moreover, when the Zeeman term predominantly lies in the $x$-$y$ plane [second row of Fig.~{\ref{fig3}}(a)], there is a clear imbalance between the two branches, and their dispersion can be fitted into Eq.~(\ref{outDispersion}). Then JDE is expected. However, for regions where either the interlayer hybridization or Zeeman terms dominates [first row of Fig.~{\ref{fig3}}(a)], the system does not develop superconducting correlation and the ABS is absent, resulting in a vanishing JDE. The $\eta$ vs $m_y$ and $m_z$ are plotted in Figs.~{\ref{fig3}}(c) and~\ref{fig3}(d), respectively. It is evident that when $m_z/\Delta<9$ is small, the system undergoes two topological phase transitions as $m_y$ increases. JDE almost vanishes in the $N=0$ phase. In contrast, both $N=1$ and $N=2$ phases manifest finite $\eta$. When the system is deep inside the $N=2$ phase, $\eta$ almost vanishes for the Zeeman terms dominate.

In Figs.~{\ref{fig3}}(e) and~\ref{fig3}(f), we present $\eta$ and $I_\text{avg}$ as functions of $(m_y,m_z)$. Here $I_\text{avg}$ serves as an indicator of the dominance of superconducting correlation. Thus $\eta$ only has a nonzero value within the region where $I_\text{avg}\neq0$. Similar to the nanowire case, $\eta$ changes significantly along part of the phase boundary, and not all the topological region exhibits a large $\eta$. However, the largest $\eta$ resides in the topological $N=1$ phase. With combined $\eta$ and $I_{\text{avg}}$, we can identify large part of $N=1$ phase by JDE, namely large JDE is an indicator for $N=1$ topological phase. However, the identification of the phase boundary by JDE is not obvious. 

\section{Conclusion} 
We have analyzed the JDE in two representative models consisting of topological superconductors in one
and two dimensions. Our results indicate that, in general, the diode efficiency can be high but not always in the topological phase. A Josephson junction in the trivial phase can also achieve relative high diode efficiency. This suggests that the existence of MBS is not a sufficient condition for realizing a large JDE. On the other hand, a significant diode effect does coincide with the topological phases, and the distinct change in diode efficiency occurs alongside some segments of the topological phase transition boundaries. In this sense, JDE with combined $\eta$ and $I_{\text{avg}}$ can serve as an indicator for topological superconductor phase. We hope the theoretical work here could aid the identification of topological superconductivity by using JDE.

\begin{acknowledgments}
This work is supported by the National Key Research Program of China under Grant No.~2019YFA0308404, the Natural Science Foundation of China through Grants No.~12350404 and No.~12174066, the Innovation Program for Quantum Science and Technology through Grant No.~2021ZD0302600, the Science and Technology Commission of Shanghai Municipality under Grant No.~23JC1400600, Shanghai Municipal Science and Technology Major Project under Grant No.~2019SHZDZX01. 
\end{acknowledgments}

{\it Note added}: During the preparation of our manuscript, we learned of an independent work on a similar problem~\cite{cayao2024enhancing}. However, their model and conclusion are different from our results.

\end{document}